\documentclass[preprint,showpacs,preprintnumbers,amsmath,amssymb]{revtex4}

\usepackage{dcolumn}
\usepackage{bm}
\usepackage{graphicx}

\begin{document}

\title{Designed Interaction Potentials via Inverse Methods for
Self-Assembly}

\author{Mikael Rechtsman$^1$} \author{Frank Stillinger$^2$}
\author{Salvatore Torquato$^{2,3}$} 
\affiliation{$^1$Department of
Physics, Princeton University, Princeton, NJ 08544}
\affiliation{$^2$Department of Chemistry, Princeton University,
Princeton, NJ 08544} \affiliation{$^3$Program in Applied and
Computational Mathematics and PRISM, Princeton, NJ 08544}

\date{\today}

\begin{abstract}
We formulate statistical-mechanical inverse methods in order to
determine optimized interparticle interactions that spontaneously
produce target many-particle configurations.  Motivated by advances
that give experimentalists greater and greater control over colloidal
interaction potentials, we propose and discuss two computational
algorithms that search for optimal potentials for self-assembly of a
given target configuration.  The first optimizes the potential near
the ground state and the second near the melting point.  We begin by
applying these techniques to assembling open structures in two
dimensions (square and honeycomb lattices) using only circularly
symmetric pair interaction potentials ; we demonstrate that the
algorithms do indeed cause self-assembly of the target lattice.  Our
approach is distinguished from previous work in that we consider (i)
lattice sums, (ii) mechanical stability (phonon spectra), and (iii)
annealed Monte Carlo simulations.  We also devise circularly symmetric
potentials that yield chain-like structures as well as systems of
clusters.

\end{abstract}

\maketitle

\section{Introduction}

``Self-assembly" of atomic, molecular and supramolecular systems is a
topic that has been receiving a great deal of attention of late.
Roughly speaking, it is the phenomenon of system components arranging
themselves via their mutual interaction to form a larger functional
unit.  Examples are plentiful; in biology, they include but are not
limited to the spontaneous formation of the DNA double helix from two
complementary oligonucleotide chains, the formation of lipid bilayers
as membranes, and spontaneous protein folding into the native,
functional state.  On the other hand, self-assembly can be employed in
the synthesis of nanostructures as an alternative to nanolithography.
For example, Whitesides \cite{whitesides1} has demonstrated that
complex two-dimensional structure can emerge in organic
molecules placed on an inorganic surface.  This is a natural system for
studying self-assembly in two dimensions.  Jenekhe and Chen
\cite{Jen99} showed self-assembly of block copolymers into ordered
arrays for possible use as photonic bandgap materials.  Block
copolymers are indeed natural candidates for use in photonic devices
due to the elaborate structures they can form and their multiple
dielectric constants.  Stellacci et. al. \cite{Stellacci} have shown
how gold nanowires can be assembled by functionalizing nanoparticles
with organic molecules.  Manoharan et. al. \cite{Pine03} have
demonstrated extremely robust self-assembly of unique, small clusters
of microspheres that can themselves be used for self-assembly of more
complex architectures.

This is an emerging field with a wealth of experimental data that does
not yet have a predictive theoretical basis.  Where there has been
theoretical work, it has focused on explaining the self-assembly in
systems with given interparticle interactions \cite{jagla, Kamien} or
of known macromolecular structure \cite{Kamien}.  These studies solve
the ``forward'' problem of statistical mechanics, i.e. they take the
interaction as known and solve for the structure and equilibrium
properties of the system.  In this study, we take the inverse appraoch
- given a desired many-particle configuration of the system, we search
for the optimal interaction among component particles which spontaneously
produces that target structure.


Our goal is to introduce an {\it inverse statistical-mechanical}
methodology for optimizing adjustable interactions for targeted
self-assembly.  Motivation for this comes from the plethora of recent
examples wherein materials have been designed to possess predetermined
properties.  Examples of these include novel crystal structures for
photonic band-gap applications \cite{Ho90}, materials with negative or
vanishing thermal expansion coefficients \cite{Si96,Mol}, materials
with negative Poisson ratios \cite{Xu99}, materials with optimal
transport and mechanical properties \cite{Hy01}, mesoporous solids for
applications in catalysis, separations, sensors and electronics
\cite{Fer99,Cha01}, and systems characterized by entropically driven
inverse freezing \cite{Gre00}.  Our goal is to devise methods that can
be applied to any predetermined target structure, be they amorphous or
even quasicrystalline, thus extending the traditional meaning of
self-assembly beyond that of periodic structures.

We choose colloidal systems \cite{Grier} as models for
studying self-assembly.  Colloids are ideally suited for this purpose
because interparticle interactions are tunable.  The colloid
interparticle potential, $V(r)$, can contain a hard-core term, a
charge dispersion (van der Waals) term, a dipole-dipole term
(isotropic in 2D), a screened-coulombic (Yukawa) term, and a
short-ranged attractive depletion term.  All of these have adjustable
amplitudes, and in the case of the Yukawa term, the screening length
can be adjusted by changing the salt concentration in solution.  Taken
together, these interactions form a large set of functional forms for
the interaction potential.  Although we do not limit ourselves in this
study to these interactions, we bear in mind the limits of complexity
that these interactions will allow and we try not to exceed these
bounds in searching for our optimized potentials.

The adjustable colloidal interactions discussed in the previous
paragraph are by nature isotropic.  Thus, in this study, we consider
only potentials that have this property.  Even for this relatively
simple class of potentials it isn't at all clear what are the
limitations for self-assembly.  For example, chiral structures with
specified handedness cannot be distinguished energetically from their
mirror-image counterpart.  What other structures cannot be valid
target structures?  A central question in colloidal and photonics
research is regarding whether a diamond lattice (in three dimensions)
can be self-assembled, since such a lattice of dielectric spheres has
a large photonic bandgap and would therefore be a viable material for
future photonic devices.  It is not known whether a diamond lattice
can be assembled using isotropic colloidal particles; indeed, the
bonding in diamond itself is highly directional.  

There has been recent interest in self-assembly of anisotropic
particles.  Examples of these are the so-called `patchy
particles'\cite{Glotzer} and the unique colloidal clusters of
Manoharan et. al. discussed above, which are are anisotropic simply by
virtue of their non-spherical shapes.  Although our algorithms can be
easily generalized to non-isotropic interactions, we restrict
ourselves to studying self-assembly with isotropic potentials since
this per se is a complex and subtle problem, and a very non-trivial
test bed for our optimization schemes.  Also, isotropic colloids are
easy to produce by comparison and their potential forms are
manipulated relatively easily.

A general potential energy function for a system of classically
interacting particles at positions $\{{\bf r}_i\}$ in zero external
field can be written as
\begin{equation}
\Phi(\{{\bf r}\}) = \sum_{i<j}V_2({\bf r}_i,{\bf r}_j) +
\sum_{i<j<k}V_3({\bf r}_i,{\bf r}_j,{\bf r}_k) + ...
\end{equation}
where the $V_\beta$'s are $\beta$-body potentials.  Since we only
consider systems with isotropic interactions, we write
\begin{equation}
\Phi(\{{\bf r}\}) = \sum_{i<j}V(|{\bf r}_i - {\bf r}_j|).
\end{equation}

Two necessary conditions for this to be useful in the present context are:
\begin{itemize}
\item The target lattice is energetically favored among a host of
other lattices over a significant specific area (denoted $\alpha$
henceforth) range (stable lattice sums).
\item That it have real phonon frequencies at every wavevector in the
Brillouin zone (stable phonons).
\end{itemize}

Past work on lattice self-assembly has not used both energy and
mechanical stability criteria in tandem as we do here; we consider
this to be a main strength of our approach.  Still, these conditions
are not universally sufficient for any pair interaction and lattice
structure.  However, taken together, these necessary conditions
constitute a prescription for finding pair potentials that most
robustly stabilize a given target lattice. In the first optimization
scheme (both are described further on), a pair potential is found that
maximizes the energy gap between the target lattices and its
competitors, while keeping all phonon frequencies real.  The second
scheme assumes stable lattice sums and real phonon frequencies, and
uses MD simulations to maximize the stability of the lattice near its
melting point.

For the purposes of this study, we will say that a lattice is
self-assembled if it is formed from a random configuration in a well
equilibrated, annealed NVT MC simulation.  It should be emphasized
that the requirement that a given lattice self-assemble in a MC
simulation is a very strong one.  In conventional theoretical studies
of colloidal crystallization \cite{Likos}, a number of candidate
lattices are chosen and a phase diagram is drawn by comparing free
energies of the lattices to each other and the liquid state over a
range of thermodynamic parameters.  However, this procedure says
nothing of mechanical stability, or whether crystallization of the
lattice is preempted by that of another structure not considered.
These shortcomings are removed when self-assembly in an MC, from a
random initial configuration, is required.  That said, finite-size
effects and limited CPU time in an MC simulation might prevent
self-assembly of a structure that should form.

In the present paper we specialize to target structures that are
two-dimensional.  In particular we seek optimal potentials for
self-assembly of the square lattice and honeycomb lattices, the latter
being the two-dimensional analog of the diamond lattice (four
maximally separated neighbors in 3D versus three maximally separated
neighbors in 2D).  This would be the first demonstration of which we
are aware of a lattice as sparse as the honeycomb being self-assembled
in an annealed Monte Carlo simulation.  This work is an expansion on a
previous introductory note by the present authors \cite{ourPRL}.  In
this paper, a new optimization algorithm is introduced and applied.
Besides the honeycomb lattice considered in the previous work, the
triangular (as a control) and square lattices are studied.  We make
the case for a more stringent requirement for self-assembly and show
that some previous claims of lattice self-assembly in linear-ramp
potentials are flawed, which is also new to this paper. A more extensive
discussion of the problem of self-assembly in systems with isotropic
interactions is given in the conclusion section, including some
novel Monte Carlo results for colloidal clusters and colloidal chains.  In a
future paper, we will apply these inverse methods to three-dimensional
colloidal systems.  While it is certainly true that many-body behavior
is fundamentally different in 3D, our methods are easily generalizable
to higher dimensions, and we believe they will be as effective.

In the following section, we discuss past work on this topic both for
the sake of motivation and to show work upon which we have attempted
to improve.  This is followed by a section describing our optimization
schemes.  Next are sections on the triangular, square and honeycomb
lattices, with potentials for their self-assembly and details of their
applications.  We discuss the triangular lattice here as a control
case, and to give the details of our simulation procedure.  The final
section is the discussion of our results and some conclusions based on
them.

\section{Previous Work}

Weber and Stillinger \cite{weberstillinger} examined self-assembly of
a square lattice for a particular potential that included two and
three body interactions.  They found that for their potential, the
square lattice was indeed the ground state and demonstrated that it
self-assembled in a 2500 particle molecular dynamics simulation.  Our
work is motivated by this, but is different in two key ways.  The
first is that we restrict ourselves to a much smaller class of
potential functions, namely those that are two body only and
isotropic.  This should make our potentials lend themselves better to
realization in the lab.  The second is that we are searching
systematically for functional forms for $V(r)$ that stabilize open
structures, whereas in \cite{weberstillinger}, the authors postulated
a potential that seemed like it should favor the square lattice and
studied its properties.  Weber and Stillinger took the direct
approach, we take the inverse approach.  

The so-called `reverse' Monte Carlo method \cite{Laaksonen} of
Lyubartsev et. al. was devised to find interparticle potentials that
produced given liquid state pair correlation functions.  A similar
method was developed by Muller-Plathe\cite{Mueller-Plathe} using
simplex optimization.  Although these are inverse techniques, they are
fundamentally different from our methods here for two reasons.  The
first reason is that the pair correlation function contains limited
information about an N-particle configuration.  Our method produces
assembly of a given configuration.  The second reason is that these
techniques fundamentally deal with liquids and so do not apply to
self-assembly as it is commonly defined.

\begin{figure}
\includegraphics[scale=0.35,clip,viewport=0pt 0pt 717pt 535pt]{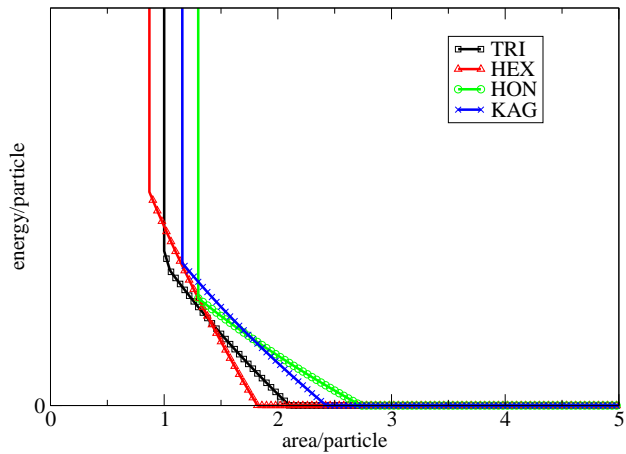}
\caption{\label{fig:jagla_ls} (Color Online) Lattice sums for the linear-ramp
potential with $r_1=1.45$.  Square wins out for specific area $\alpha=1$ to
$\sim\alpha=1.4$. }  \vspace{0.2in}
\end{figure}

\begin{figure}
\includegraphics[scale=0.35,clip,viewport=0pt 0pt 717pt 535pt]{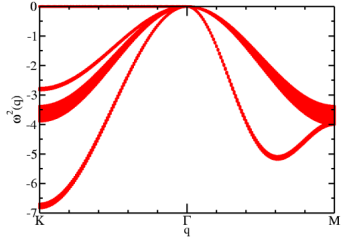}
\caption{\label{fig:jagla_pho} (Color Online) Phonon spectra for
specific area $\alpha=1.0$ to $\alpha=1.4$ for the square lattice in
the linear ramp potential with $r_1=1.45$.  Bands form as a result of
the variation in $\alpha$.  Over this entire range of $\alpha$, all
frequencies are imaginary, which indicates mechanical instability in
the lattice.  Over this density range, the square lattice is clearly
not the ground state.}
\vspace{0.2in}
\end{figure}

\begin{figure}
\includegraphics[scale=0.45,clip,viewport=0pt 0pt 430pt
430pt]{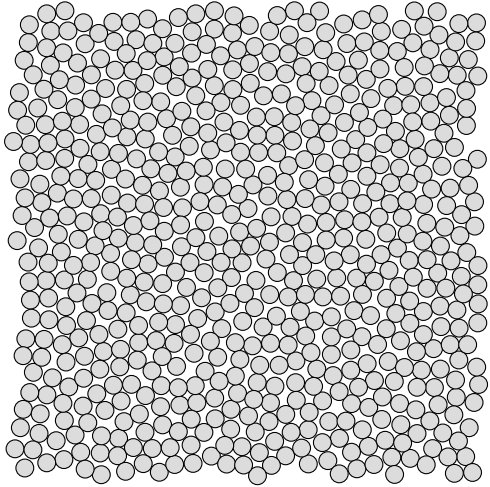}
\caption{\label{fig:jagla_mc} 625 particle MC results for the
linear-ramp potential.  Annealed from $k_BT=1.1$ to $k_BT=0.02$.}
\end{figure}

\begin{figure}
\includegraphics[scale=0.45, width=2.6in, height=2.6in]{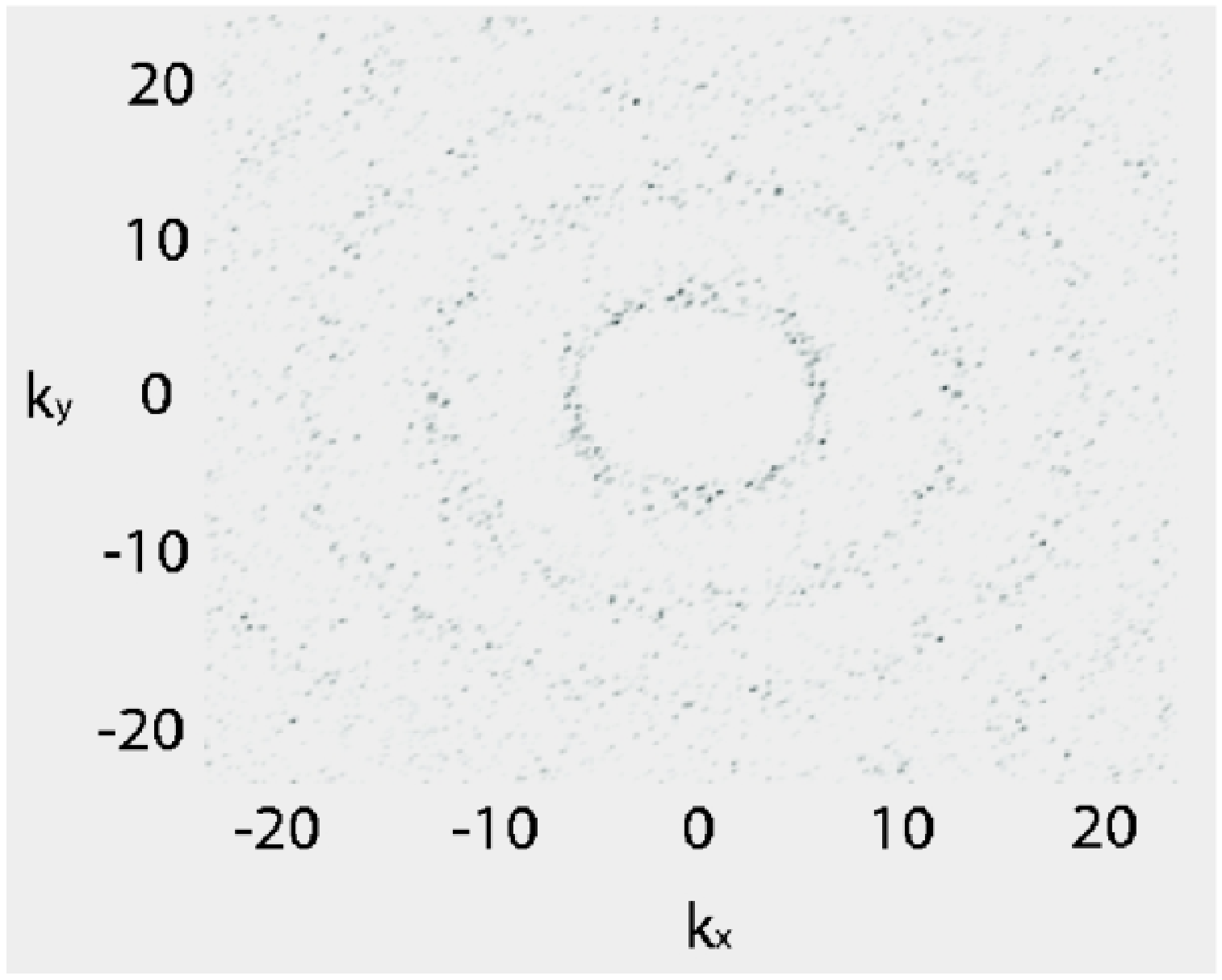}
\caption{\label{fig:jagla_sf} S(k) for configuration in figure \ref{fig:jagla_mc}.}
\end{figure}

E. A. Jagla claims in \cite{jagla} to have found an isotropic pair
potential form that stabilizes a number of structures (including the
honeycomb lattice) called the `linear-ramp' potential, which consists
of a hard-core at $r=1$ plus a linear tail going to zero at a distance
$r_1>1$.  A phase diagram is drawn in that paper indicating the
stability of the structures he lists for different pressures and
values of $r_1$.  However, the structures which he gives as stable at
a number of points in his phase diagram do not meet both of the
necessary conditions for self-assembly that we describe above, and
they do not self-assemble in annealed MC simulations.  As a
demonstration of this, we choose $r_1=1.45$, which the phase diagram
indicates should yield the square lattice for certain pressure values.
For this value of $r_1$ we find the range in specific area, $\alpha$,
for which the square beats out the other three lattices (see
Fig. \ref{fig:jagla_ls}), and find the phonon spectra (see Fig.
\ref{fig:jagla_pho}) over that range.  Phonon spectra were calculated
in the standard way by diagonalizing the dynamical matrix for a very
fine grid of k-points in the Brillouin zone (a detailed explanation of
this is given in \cite{Ashcroft}).  Any lattice at a given specific
area/volume that has an imaginary phonon frequency at any wavevector
is necessarily mechanically unstable.  An NVT MC simulation of 625
particles annealed from $k_BT=1.5$ to $k_BT=0.05$ is shown at
$\alpha=1.38$ in Fig. \ref{fig:jagla_mc}.  Although there are pockets
of square lattice present, it is clear that the lattice has not
assembled, and there is no long range order.  This can be seen from
the plot of the structure factor, $S(k)$, given in Fig.
\ref{fig:jagla_sf}.  One of two things has happened here.  Either the
system has become a glass, or it has no freezing transition.  This
shows that comparing the energies of a number of lattices (as
suggested in \cite{jagla}) cannot alone give certainty of the ground
state.

\section{The Optimization Schemes}

A central feature of our approach to the inverse problem is the design
of computational algorithms that search for and optimize a functional
form for $V(r)$ that leads to self-assembly of a given target structure.
The direct (non-inverse) version of this is the problem of the first
order freezing transition, and has been studied analytically and
numerically using, for example, classical density functional methods
\cite{ramakrishnan}.  

Optimizing a pair potential, $V(r)$, for self-assembly means choosing
a family of functions $V(r; \{a_0...a_n\})$, parameterized by the
$a_i$'s, and then finding the values of the parameters that lead to
the most robust and defect-free self-assembly of the target lattice,
at a given specific volume $\alpha$ (or specific area in 2D).  We
must be careful to choose the parameters such that an overall
rescaling of the potential is not possible, and we keep each parameter
within a prespecified range, $[a_i^{min}, a_i^{max}]$.  The choice of
parameterization and initial parameter values is important: we make
educated guesses based on the coordination numbers of lattices close
in structure to the target lattice.  Optimization can be carried out
either at zero temperature or near melting.  \\

\subsection{T=0 Optimization Scheme}

Once the parameterization and initial parameters are chosen, we
perform a simulated annealing optimization to maximize the difference
in lattice energy per particle, $\epsilon$, between the target lattice
and its closest energetic competitor among the principle lattices in
the system dimension (e.g. in 2D, among triangular, square, honeycomb
and Kagom\'e).  This procedure is called the `zero-temperature' scheme
because it seeks to minimize the difference in lattice potential
energies, rather than free energies; it is a search for stable ground
states.  Formally, if $\epsilon(\alpha)$ is the energy per particle at
specific volume $\alpha$, we take as our objective function
\begin{equation}
\Theta_1 = \max_{j}[\min_{\alpha \in [\alpha_{min}, \alpha_{max}]}
\epsilon^{T}(\alpha)- \min_{\alpha \in [\alpha_{min}, \alpha_{max}]}
\epsilon^{j}(\alpha)].
\end{equation} 

Here, $T$ refers to the target lattice, $j$ enumerates the competitor
lattices and $ [\alpha_{min}, \alpha_{max}]$ is a specific volume
range, within which the target $\alpha$ lies.  The simulated annealing
is performed in $a_i$ parameter space, searching for a potential that
minimizes $\Theta_1$.  This alone is not sufficient; we must also
guarantee the mechanical stability of the lattice.  This is done by
making sure that at the target $\alpha$, the given potential is such
that every phonon mode in the Brillouin zone is real.  In practise,
this is done by constraining the lowest eigenvalue of the dynamical
matrix (frequency squared) to be positive, and the lowest curvature
eigenvalue of the softest acoustic phonon mode to be greater than some
positive cutoff value.  While this does not necessarily imply that all
frequencies will be real, it is usually sufficient, and in any case
the frequency of every mode can be calculated post facto.

In this scheme, we make the assumption that the greater the difference
in lattice energy per particle, $\epsilon$, (over a range of
$\alpha$'s around the target $\alpha$) of the target lattice and its
principle competitors, the greater will be the target's tendency to
assemble.  While this is not by any means a rigorous statement, it
seems to make intuitive sense - the greater the energy difference, the
less the tendency to get frustrated at the freezing point between two
lattices; the annealing should find the deeper energy minimum.

It is possible that another structure will preempt the target lattice
(freeze at a higher temperature), even if the optimization proceeds
perfectly.  The MC will then get `stuck'; the simulation will never go
to its ground state because it is caught in a strongly metastable
state.  Presumeably, however, a colloidal system with the same
interaction potential would undergo a structural phase transition to
the its ground state as the temperature was lowered.  Our MC
simulations did indeed get stuck in slightly defected configurations
very close to the desired lattice.  To check that these structures
weren't inherently more stable than the target, we always confirmed
that the defects caused the system to have higher energy than that of
the lattice.

The main disadvantage of this optimization procedure is that it is very
specific to simple lattices, and is not naturally generalized to more
complicated structures.  Indeed, the CPU time required for the
optimization grows as the cube of the number of basis elements in the
lattice, so optimizing for complex structures quickly becomes
intractable.  Nonperiodic structures (e.g. quasicrystals) are thus
impossible for this scheme.  

This optimization scheme is competitor-based; we favor the target by
energetically disfavoring other lattices.  However, this does not
preclude other structures, periodic or otherwise from being lower in
energy than the target.  This is an inherent limitation of this
technique.  The next scheme, however, does not suffer this
shortcoming.

\subsection{`Near melting' optimization scheme}

In this procedure, we first make sure that the initial potential
satisfies our two stated necessary conditions for self-assembly with
the initial parameter values, namely that the target lattice is
energetically favored over the others over a wide $\alpha$-range, and
that at our chosen $\alpha$, all phonon modes are real.  We then feed
this family of functions to the algorithm, and optimize it for
self-assembly at a temperature near (but below) the lattice's melting
point by suppressing nucleation of the liquid phase in MD simulations.

We first find the melting temperature of the system by running an NVE
(canonical ensemble) molecular dynamics simulation (MD) on a system of
particles, in the target configuration, at incrementally increasing
temperatures (mean square velocity).  We then run the MD repeatedly at
80-95\% of the melting temperature (the temperature is chosen such
that phase-transition fluctuations do not render the calculations
inconsistent), each time calculating the Lindemann parameter, defined
by


\begin{equation}
\Theta_2 = \sqrt{\frac{1}{N}\sum_i ({\bf r}_i - {\bf r}_i^{(0)})^2 -
\left( \frac{1}{N}\sum_i ({\bf r}_i-{\bf r}_i^{(0)})\right)^2},
\end{equation}

where ${\bf r}_i$ is the position of the $i^{th}$ particle after an
appropriate amount of simulation time, ${\bf r}_i^{(0)}$ is its
initial position, and $N$ is the number of particles.  $\Theta_2$ is
then taken as the objective function for a simulated annealing
calculation, and those parameters, $a_i$, are found such that
$\Theta_2$ is minimized.  It should be noted that in order to get a
reproducible value of $\Theta_2$, it must be averaged over a number of
MD runs.

We choose to minimize the Lindemann parameter because it gives some
quantitative measure of the degree of liquid nucleation or structural
phase transition setting in near the melting point.  Presumeably, the
more these effects are suppressed, the more robustly the potential
favors the given target structure.  The algorithm will by its nature
disfavor potentials that violate either of our two necessary
conditions for self-assembly.  It is inevitable that over the course
of the optimization the melting temperature of the potential will be
changed; it could be that at that point, the system will no longer be
near the phase coexistence regime.  This can be detected easily enough
(for example by comparing the Lindemann parameter to that which the
harmonic approximation predicts), and then the optimization can be
stopped and restarted at a higher, appropriately chosen temperature.

An important limitation of this optimization is in the tradeoff between
its consistency and its closeness in temperature to the melting point.
Due to large fluctuations near melting, getting reproducible values
for $\Theta_2$ with sufficiently small error requires larger and
larger system sizes.  So while the optimization can be carried out
well into the anharmonic regime, the optimization cannot sample true
phase coexistence, only nucleation.

The inherent bias in this scheme towards the target lattice presents a
problem for optimization.  The procedure does not distinguish between
a configuration being in a thermodynamically stable state at the given
temperature and being in a supercooled metastable state.  Just as in
a MC simulation, the MD may get `stuck'.  As a result, the target may
become strongly metastable but never thermodynamically favored.  The
only way to decrease this effect is to get closer in temperature to
the melting point, but this in turn requires more and more CPU time.

Besides the obvious advantage that this scheme incorporates
finite-temperature, anharmonic effects, it has the advantage of being
competitor-free.  Whereas in the $T=0$ scheme, competitor lattices
have to be chosen against which the target lattice competes, this
procedure ostensibly optimizes against {\it all} competition.  
It should be noted that if an initial potential with favorable lattice
sums and stable phonons for this procedure cannot be found by
trial-and-error, the zero-temperature scheme can be run first on a
given functional parameterization.  This procedure would then take
that output as its input.  This is perhaps the best way to combine the
two optimizations.


\section{The Triangular Lattice}

\begin{figure}
\includegraphics[scale=0.35,clip,viewport=0pt 0pt 717pt 535pt
]{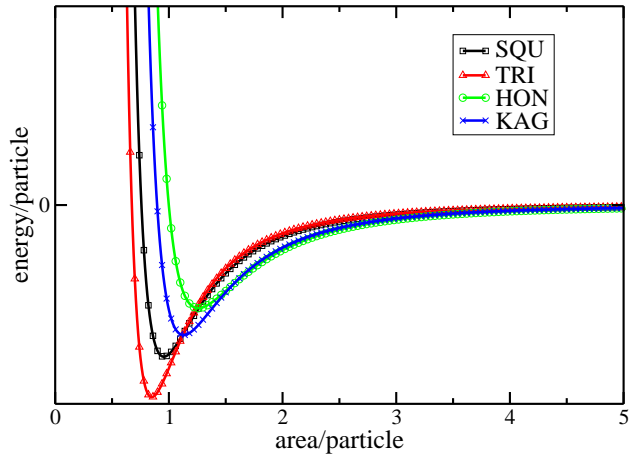}
\caption{\label{fig:tri_lat} (Color Online) Lattice sums for the LJ potential.}
\end{figure}

\begin{figure}
\includegraphics[scale=0.35,clip,viewport=0pt 0pt 717pt 535pt
]{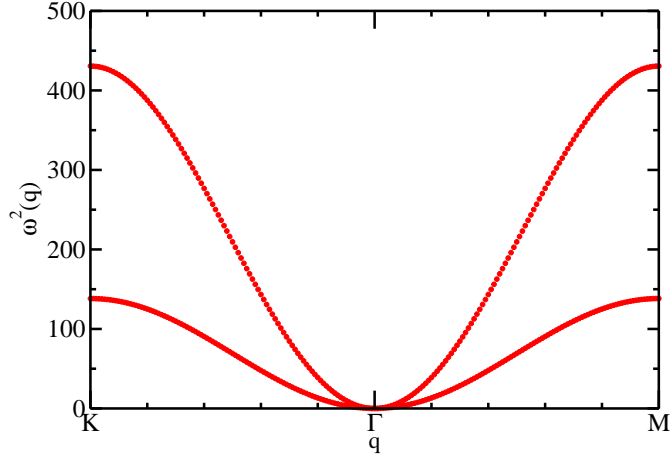}
\caption{\label{fig:tri_pho} (Color Online) Phonon spectrum for triangular with LJ
potential at $\alpha=\sqrt 3/2$. }
\vspace{0.2in}
\end{figure}

\begin{figure}
\includegraphics[scale=0.45,clip,viewport=0pt 0pt 330pt
330pt]{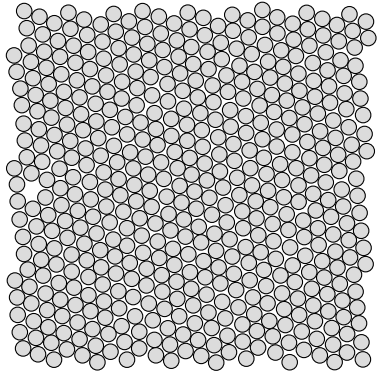}
\caption{\label{fig:tri_mc} 500 particle MC results annealed from
$k_BT=1.5$ to $k_BT = 0.2$ at $\alpha=\sqrt 3/2$.}
\end{figure}

\begin{figure}
\includegraphics[width=2.6in, height=2.6in]{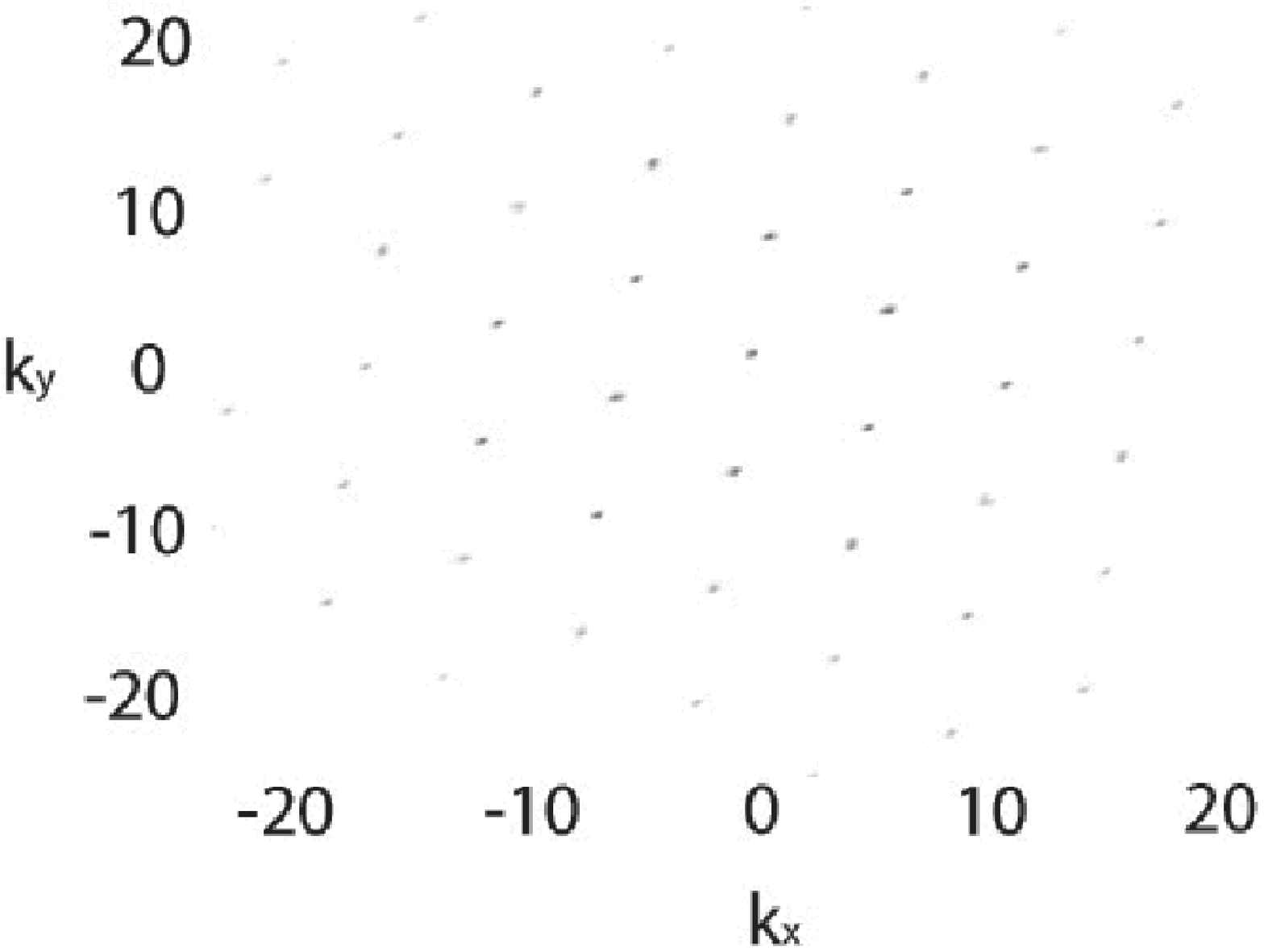}
\caption{\label{fig:tri_sf} S(k) for the configuration shown in Fig.
\ref{fig:tri_mc}.}
\end{figure}

A very well studied interparticle potential that robustly stabilizes
the triangular lattice is the LJ\cite{colloidsinteresting}, given in a
form rescaled from its traditional definition,
$$
V(r) = \frac{1}{r^{12}} - \frac{2}{r^6}.
$$ We discuss this potential here as a control.  We have employed it
in a 500 particle NVT ensemble, annealing it down from $k_BT = 3.0$ to
$k_BT = 0.2$ (allowing sufficient equilibration time at each
temperature step), with average specific area given by the triangular
lattice area when the nearest neighbor is at unity, namely
$\sqrt{3}/2$.  Lattice sums, shown in Fig. \ref{fig:tri_lat},
demonstrate that energetically, the triangular does beat the square,
honeycomb, and Kagom\'e lattices over a wide range of $\alpha$'s
(actually globally in this case).  Fig. \ref{fig:tri_pho} shows that
all phonon frequencies are indeed real.  The two branches of course
represent the longitudal and transverse acoustic modes of oscillation.
Clearly the LJ potential meets our two necessary conditions, that it
be energetically favored over the other lattices and that it have real
phonon frequencies.  Fig. \ref{fig:tri_mc} shows that it does indeed
self-assemble into the triangular lattice.  The structure factor,
$S(k)$, shown in Fig. \ref{fig:tri_sf} shows conclusively the
existence of long-range order here.  For a different target lattice,
we would have defined a family of potentials of which the LJ was one,
run the optimization program, and then performed the MC self-assembly
calculation.  We do not run the optimization for the LJ here because
it is a relatively simple potential, and we wish to use it simply as a
reference.  Note that for all MC simulations in this study we use an
NVT ensemble with periodic boundary conditions, adjusting the MC
maximum step fraction such that 30\% acceptance is maintained
throughout the simulation (for maximal ergodicity)\cite{camp}, and
anneal through the freezing transition towards $T=0$.

\section{The Square Lattice}

Finding a stabilizing potential for the square lattice is in some
sense a more straightforward task than for the honeycomb lattice since
the neighbor distances are different from those of the triangular.
Quandt and Teter accidentally came across a pair interaction that
stabilized the square lattice when examining quasiperiodic structures
in 2D systems \cite{quandt:8586}.  Qualitatively very similar to the
optimal $V(r)$ that we derive below, their square lattice potential
satisfies our necessary conditions, as expected.  Their potential
gives a very soft phonon branch, causing the crystal to be very
sensitive to perturbations - the potential we derive below improves on
this.  Here we use the square lattice as a simple illustration and a
test case of our methods.  In finding an initial potential, we choose
to start with the LJ potential.  Consider an $\alpha$ for which the
nearest neighbor distance is unity (for the square lattice, this is
itself unity, i.e. $\alpha=1$), then for the triangular lattice, the
next nearest neighbor is at $r=\sqrt 3$ and for the square lattice it
is at $r=\sqrt 2$.  Thus, we desire to find a potential that is
positive at $r=\sqrt 3$ but negative at $r=\sqrt 2$.  Consider a LJ
potential with an added Gaussian centered at $\sqrt 3$ which has a low
enough width so that $V(\sqrt 2)$ is negative, and has a great enough
amplitude so that $V(\sqrt 3)$ is positive.  This would do the job of
favoring the square lattice second neighbor while excluding the
triangular lattice one.  Still, the amplitude and width must be chosen
such that our necessary stability conditions are met.  The trade-off
here is clear: with an amplitude too low, the square lattice will not
be energetically favored, and with an amplitude too high, the lattice
will not be mechanically stable (the phonon frequencies will not be
everywhere real).  We have found such a potential, namely,
\begin{equation}
V_{SQU}(r) = \frac{1}{r^{12}} - \frac{2}{r^6} + 0.7\exp[-25(r-\sqrt 3)^2].\label{vsqu}
\end{equation}

\begin{figure}
\includegraphics[scale=0.35,clip,viewport=0pt 0pt 717pt 535pt
]{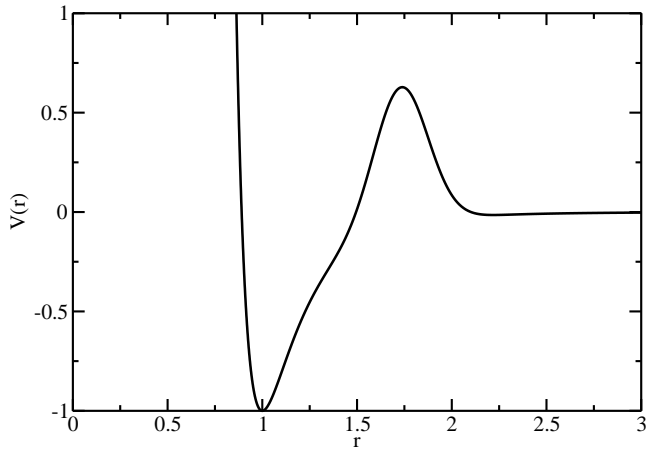}
\caption{\label{fig:squ_pot} $V_{SQU}$, as given in (\ref{vsqu}).}
\end{figure}

\begin{figure}
\includegraphics[scale=0.35,clip,viewport=0pt 0pt 717pt 535pt
]{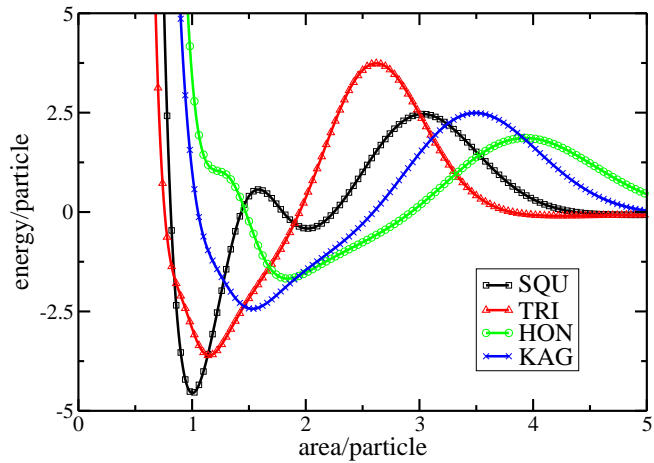}
\caption{\label{fig:squ_lat} (Color Online) Lattice sums for $V_{SQU}$.}
\end{figure}

\begin{figure}
\includegraphics[scale=0.35,clip,viewport=0pt 0pt 717pt 535pt
]{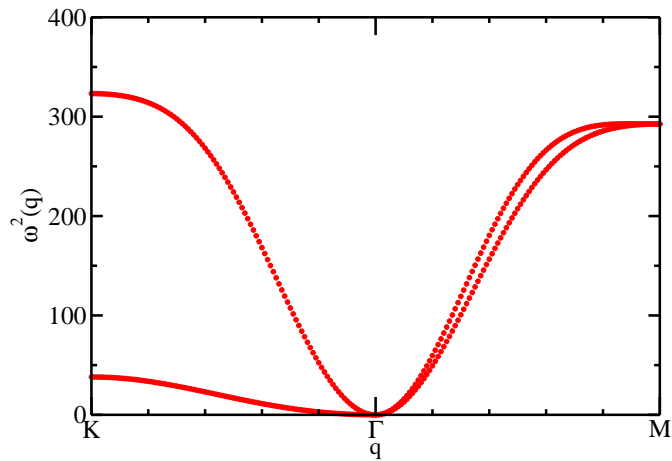}
\caption{\label{fig:squ_pho} (Color Online) Phonon spectrum for square lattice with potential
$V_{SQU}$ at $\alpha = 1.0$. }
\vspace{0.2in}
\end{figure}

This potential is plotted in Fig. \ref{fig:squ_pot}.  The lattice sums
for $V_{SQU}(r)$ are given in Fig. \ref{fig:squ_lat} and its phonon
spectrum is shown in Fig. \ref{fig:squ_pho}.  Clearly the square
lattice is energetically favored and mechanically stable.  The Maxwell
double tangent construction applied to the lattice sums gives a range
of stability in pressure of $0$ through $23.3$, and in specific area
of approximately $0.85$ through $1.0$.  We parameterize this potential
as follows:
\begin{equation}
V_{SQU}(r; a_0, a_1, a_2) = \frac{1}{r^{12}} - \frac{2}{r^6} +
a_0\exp[-a_1(r-a_2)^2].
\end{equation}  

We then choose bounds for the parameters, somewhat arbitrarily (such
that the final potential still resembles the initial guess).  We ran
the near-melting and the zero-temperature optimization schemes.  The
near-melting optimization produced the potential
\begin{equation}
V_{SQU}(r) = \frac{1}{r^{12}} - \frac{2}{r^6} +
0.828\exp[-26.5(r-1.79)^2], \label{squ-pot-2}
\end{equation} 
and the zero-temperature optimization produced
\begin{equation}
V_{SQU}(r) = \frac{1}{r^{12}} - \frac{2}{r^6} +
0.672\exp[-42.242(r-1.8248)^2], \label{squ-pot-1}
\end{equation}

\begin{figure}
\includegraphics[scale=0.45,clip,viewport=0pt 0pt 340pt
340pt]{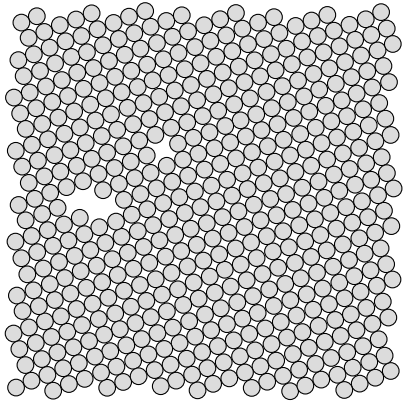}
\caption{\label{fig:squ_mc} 484 particle MC results annealed from
$k_BT=1.0$ to $k_BT = 0.1$ at $\alpha = 1$. }
\end{figure}

\begin{figure}
\includegraphics[width=2.6in, height=2.6in]{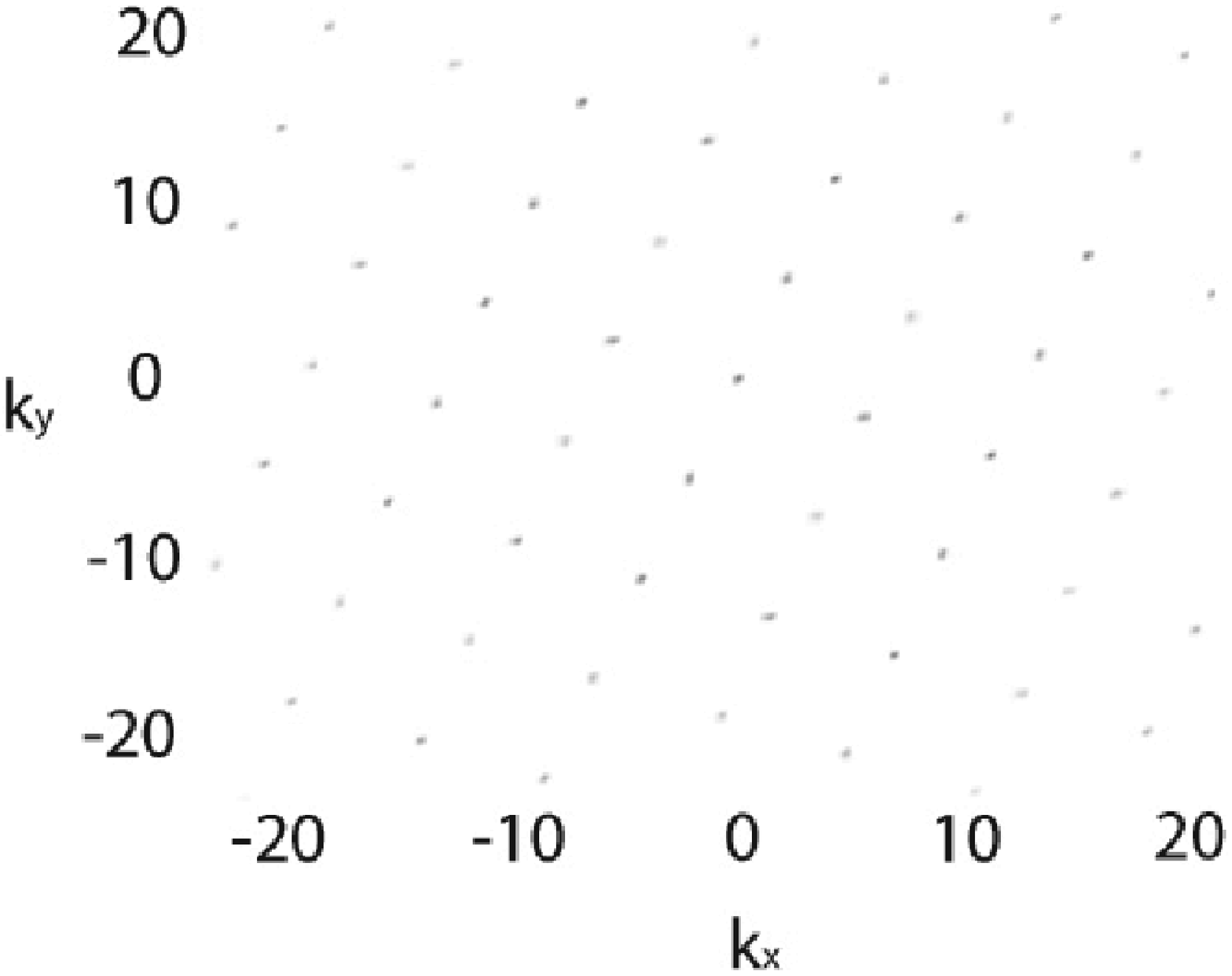}
\caption{\label{fig:squ_sf} $S(k)$ for configuration given in figure
\ref{fig:squ_mc}.}
\end{figure}

The square lattice potentials are run in 484-particle MC calculations,
annealed to $k_BT = 0$ from $k_BT = 1.0$, at $\alpha=1.0$. We find
that the potentials from both optimization schemes cause square
lattice self-assembly, as is evidenced in Fig. \ref{fig:squ_mc} (the
MC results), and in Fig.  \ref{fig:squ_sf}, the structure factor,
which shows the presence of long-range order.  The results shown are
for the near-melting optimization, but we obtained essentially the
same results for the zero-temperature optimization.  Thus, we have
`solved' the inverse problem for the case of the square lattice, or at
least we have found two working solutions.

\section{The Honeycomb Lattice}


We base our choice for the parameterization of the honeycomb pair
potential on the fact that the honeycomb is a sublattice of the
triangular, sharing the same neighbor distances.  The first and second
coordination numbers are (3,6) and (6,6) for the honeycomb and
triangular lattices respectively.  We therefore choose a potential
that is positive at what we intend to be the nearest neighbor
distance.  For the sake of mechanical stability (real phonon
frequencies), we put a potential `well' at that distance, in the form
of a 12-10 Lennard-Jones potential.  Including an exponential
repulsive term, we first parameterize the potential as follows:
$$
V(r; a_1, a_2) = \frac{5}{r^{12}} - \frac{6}{r^{10}} +  a_1\exp[-a_2r].
$$ 

Phonon frequencies could not all be made real using this
parameterization, and thus it was deemed to be insufficient.  As a
result, we add to the parameterization an attractive Gaussian of set
depth and variance, meant to `brace' the second neighbor:
\begin{equation}
V_{HON}(r; a_0, a_1, a_2, a_3) = \frac{5}{r^{12}} -
\frac{a_0}{r^{10}} + a_1\exp[-a_2r] - 0.4\exp[-40(r-a_3)^2].
\end{equation}
Note that here we are now allowing the coefficient of the
$1/r^{10}$ term to vary in the optimization.  After some
encouraging phonon spectra, lattice sums and annealing results with a
number of different parameter value inputs, we concluded that this was
a sufficiently (but not overly) complex functional form on which to
perform the optimization.  The targeted specific area is $\alpha =
1.45$.

\begin{figure}
\includegraphics[scale=0.4,clip,viewport=0pt 0pt 450pt
450pt]{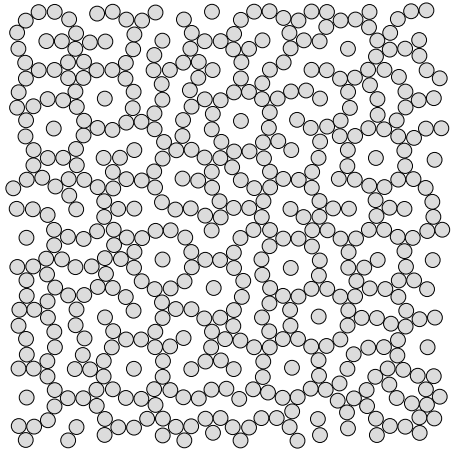}
\caption{\label{fig:hon_mc_bad} 500-particle annealed
MC results, for potential with parameters displaced from initial
guess. $\alpha = 1.45$.  }
\end{figure}

The initial values for the parameters were chosen to be $a_0=6.5,\
a_1=18.5, \ a_2=2.45,$ and $a_3=1.83$.  For comparison to optimized
results, a 500-particle annealed MC simulation was run using these
parameters, the result of which is shown in Fig.
\ref{fig:hon_mc_bad}.  It is clearly nowhere resembling a honeycomb
lattice configuration.

\begin{figure}
\includegraphics[scale=0.35,clip,viewport=0pt 0pt 717pt 535pt
]{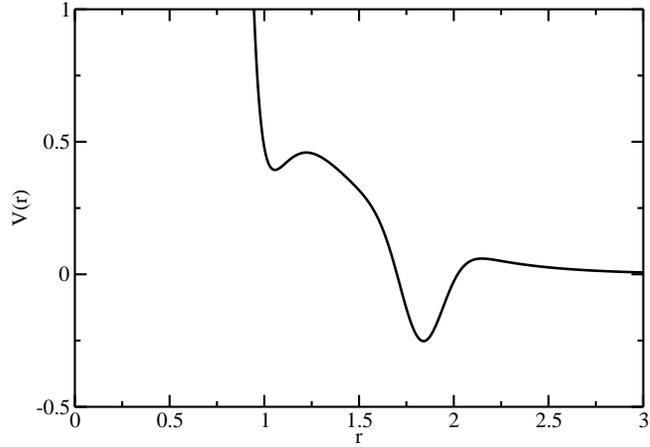}
\caption{\label{fig:hon_pot} $V_{HON}$, given in (\ref{hon-pot-2}).}
\end{figure}

\begin{figure}
\includegraphics[scale=0.35,angle=270,clip,viewport=0pt 0pt 717pt 735pt
]{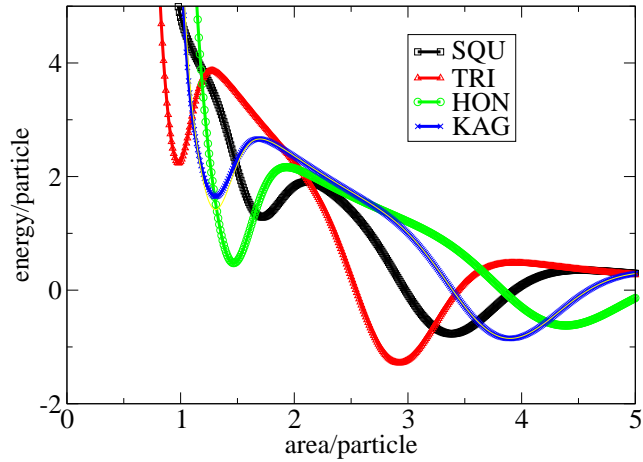}
\caption{\label{fig:hon_lat} (Color Online) Lattice sums for $V_{HON}$.  Note that we choose an $\alpha$ at which pressure is positive, i.e. slightly lower than the local energy minimum.}
\end{figure}

\begin{figure}
\includegraphics[scale=0.35,clip,viewport=0pt 0pt 717pt 500pt
]{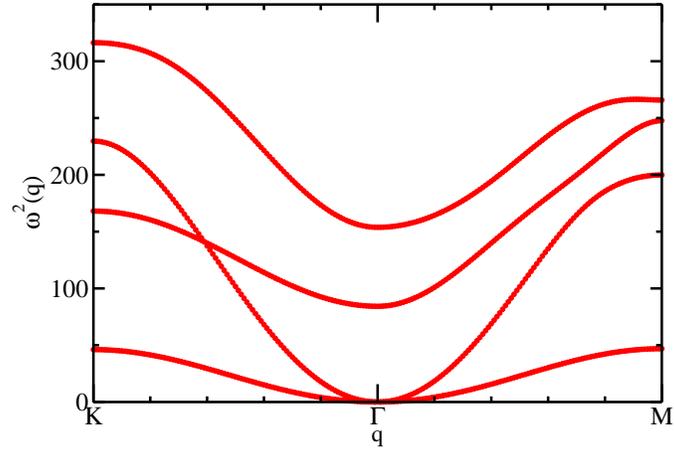}
\caption{\label{fig:hon_pho} (Color Online) Phonon spectrum for honeycomb lattice with
potential $V_{HON}$ at $\alpha = 1.45$. }
\vspace{0.2in}
\end{figure}

Both optimization schemes were carried out on this parameterization of
the potential.  Here, we show all results for the near-melting
scheme and simply state the results for the zero-temperature scheme.
The near-melting algorithm produced the following potential:
\begin{equation}
V_{HON}(r) = \frac{5}{r^{12}} -
\frac{5.89}{r^{10}} + 17.9\exp[-2.49r] - 0.4\exp[-40(r-1.823)^2].\label{hon-pot-2}
\end{equation}
This function is plotted in Fig. \ref{fig:hon_pot}.  The lattice sums
and phonon spectrum are given in Figs. \ref{fig:hon_lat} and
\ref{fig:hon_pho} respectively.  Notice that in the region of
stability of the honeycomb lattice the pressure (i.e.,
$-\frac{\partial e}{\partial \alpha}$, where $e$ is the energy per
particle at $T=0$) would have to be positive in order to ensure
thermodynamic stability.  The reader should note, however, that in
principle, it is always possible to append to a constructed pair
interaction a weak long-ranged attractive component
(Kac-Uhlenbeck-Hemmer potential\cite{Kac}); the corresponding
influence on the lattice sums is to subtract a contribution
proportional to the number density, thus lowering the corresponding
lattice sum toward a positive pressure regime.  As they are, the
lattice sums give a range of stability in pressure of $1.2$ through
$3.8$, and in specific area of approximately $1.42$ through $1.48$.

\begin{figure}
\includegraphics[scale=0.45]{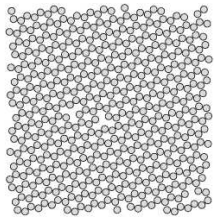}
\caption{\label{fig:hon_mc} 500-particle MC results annealed from
$k_BT = 0.5$ to $k_BT = 0.05$ at $\alpha = 1.45$ for potential in Fig. \ref{fig:hon_pot}. }
\end{figure}

\begin{figure}
\includegraphics[width=2.6in, height=2.6in]{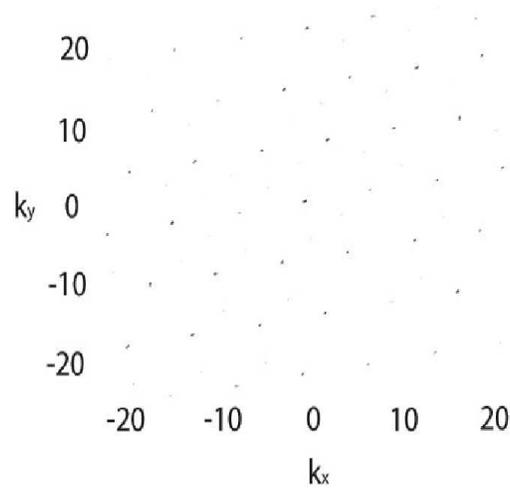}
\caption{\label{fig:hon_sf} $S(k)$ for configuration given in Fig.
\ref{fig:hon_mc}.}
\end{figure}

The 500-particle annealed MC simulation for this potential is shown in
Fig. \ref{fig:hon_mc}.  The structure factor, $S(k)$, for this
configuration is shown in Fig. \ref{fig:hon_sf}, and it indicates the
presence of long-range order.  Self-assembly has been achieved -
although there are clearly defects, these were simply ``frozen in''
during annealing.  Their presence {\it costs energy}, indicating that
the defective structure is not the true ground state, as expected.
The zero temperature scheme produced the potential
\begin{equation}
V_{HON}(r) = \frac{5}{r^{12}} -
\frac{6.50}{r^{10}} + 18.19\exp[-2.21r] - 0.4\exp[-40(r-1.755)^2].\label{hon-pot-1}
\end{equation}

Similarly to the square lattice, the zero-temperature scheme produced
a honeycomb structure with long-range order, albeit with more defects
(11 vacancies and 2 interstitials, compared to 3 vacancies and 0
interstitials for the function given in (\ref{hon-pot-2})).

\section{Conclusions and Discussion}

In sum, we have introduced and demonstrated two optimization schemes
for lattice self-assembly in two dimensions, each producing optimized
pair potentials for the square and honeycomb lattices.  The schemes
are directly generalizable to three dimensions and to more complicated
structures.  Future work will do exactly this, testing whether schemes
that work well for single component systems in two dimensions have
wider applicability.

Although we have found potentials that have as their ground
states the honeycomb and square lattices at particular $\alpha$'s,
these functional forms for $V(r)$ are by no means unique.  In future
work, we will try to optimize for `robustness' in self-assembly.  In
particular, we would like to find potentials that not only cause
self-assembly of a system of particles into a desired target
structure, but that make the structure minimally sensitive to
perturbations in density, pressure, and chemical potential, as well as
to small changes in functional form of $V(r)$.  This is extremely
important if these potentials are to be implemented experimentally for
two reasons.  The first is that there is of course some experimental
error in tuning the parameters in the potential, and these small
uncertainties should not prevent self-assembly.  The second is that we
may wish to use experimental interactions to approximate optimal
solutions with different parameterizations, and there will be some
error associated with this fit.  The potentials for the square and
honeycomb lattices found in this work can indeed be called robust.
For the square lattice potential, there is a wide range of parameter
values around our optimal solutions that yield favorable lattice sums,
real phonon frequencies, and produce near defect-free self-assembly.
The important features of this potential are a strong initial
repulsion (representing a near hard-core interaction), an attractive
well at distance $\sqrt \alpha$ as well as a positive maximum (we used
a gaussian) at or around $\sqrt{3\alpha}$.  Not any functional form
with these features will necessarily work; but we have found that
perturbations around the potentials given above (relations
\ref{squ-pot-2} and \ref{squ-pot-1}) that preserve these features do
indeed cause square lattice self-assembly.  The same can be said of
the potentials derived for the honeycomb lattice (relations
\ref{hon-pot-2} and \ref{hon-pot-1}), except of course with different
features.  These features are the strong initial repulsion, the
positivity of the first minimum and the negativity of a second
minimum, where the minima are at distance ratio $\sim \sqrt 3$.  The
chosen $\alpha$ puts the first minimum at, or near, the honeycomb
nearest neighbor.

It is a natural question to ask whether available colloidal
interactions can be made to fit our optimized $V(r)$'s.  Although
obviously they cannot match these functions exactly, they can indeed
form a good approximation.  For example, our optimized honeycomb
potential has a strong initial repulsion, followed by a short
attraction, a steep repulsion and then another attraction.  By
adjusting relative amplitudes, this functional form can be
approximated by a hard-core, a dispersion interaction, a repulsive
dipole-dipole interaction and an attractive depletion.  There is
indeed hope for using realizable interactions to form open structures
in 2D colloidal systems.

Extensive attempts were made to find a potential that stabilizes the
Kagom\'e lattice but none were thoroughly successful.  A potential was
found that satisfied the necessary conditions, and the optimization
was run.  Although the MC run gave a lattice with long-range order,
the interstitials were somewhat randomly placed.  This is because
there exists another 2D lattice with 4 coordination (as the Kagom\'e
has)\cite{kagomealt}, and it and the Kagom\'e are nearly
indistinguishible in energy for almost any LJ-based potential we used.
Because of this closeness in energy, the interstitial sites themselves
formed a weakly interacting lattice gas.

The problem of Kagom\'e lattice self-assembly goes beyond the
competition with one other lattice, however.  The Kagom\'e has the
property that it is in many ways an intermediate between the
triangular and honeycomb lattices: its density is in between the two;
the first three coordination numbers are (4,6,4) as opposed to (6,6,6)
in the triangular and (3,6,3) in the honeycomb lattice.  So if the
potential is significant for only the first three neighbors (as ours
have been) then, rigorously, the Kagom\'e cannot energetically beat
the honeycomb and triangular lattices over all densities.
Furthermore, there is an extremely delicate energetic balance between
a stable Kagom\'e and a phase separation into the triangular and
honeycomb lattices.  We believe that for these reasons our
optimizations have been unable to cause self-assembly of the Kagom\'e.

The defects that were observed in the self-assembled triangular and
honeycomb lattice do not disturb the crystal structure; this is to say
that particles could be simply inserted at the defect points and a
perfect lattice would emerge.  There are two possible reasons for the
defects: first, that to have some number of them is energetically
favorable, and thus the ground state configuration is not the perfect
lattice; second, that the slow dynamics of the MC simulation at low
temperatures prevent the defects from being removed in a realistic
amount of CPU time.  A look at the energetics shows that the second
explanation is the right one.  Over a small range of $\alpha$ around
the simulation density, the perfect lattice has lower energy than that
of the structure obtained in the simulation.  Perhaps simulations in
the $\mu$VT ensemble (grand canonical) would allow for more general
phase space sampling and thus demonstrate that the gaps are
spontaneously filled.

\begin{figure}
\includegraphics[scale=0.35]{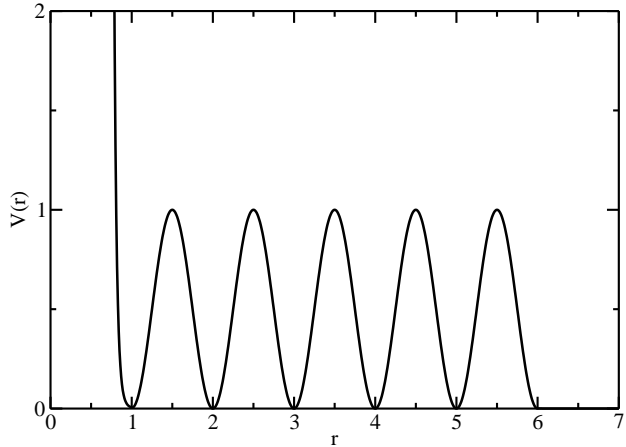}
\caption{\label{fig:5-finger} 5-finger potential.}
\end{figure}

\begin{figure}
\includegraphics[scale=0.45,clip,viewport=0pt 0pt 500pt 450pt
]{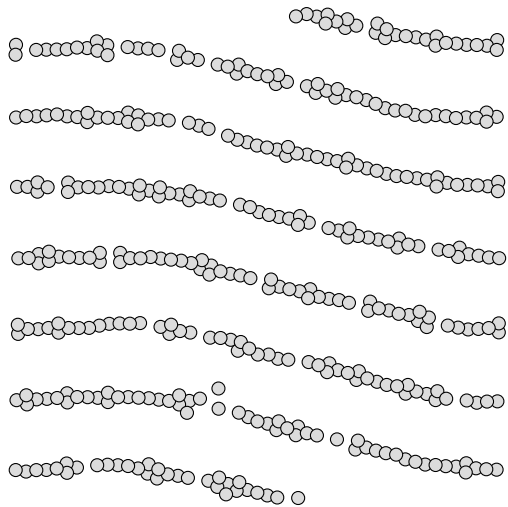}

\caption{\label{fig:5-finger-mc} 384 particle MC results for 5-finger
potential annealed to $k_BT=0.1$.}
\end{figure}

\begin{figure}
\includegraphics[scale=0.35,clip,viewport=0pt 0pt 717pt 535pt
]{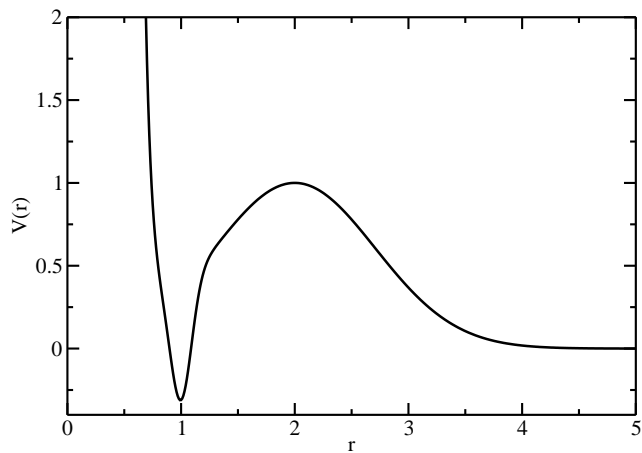}
\caption{\label{fig:simplex} Simplex potential.}
\end{figure}

\begin{figure}
\includegraphics[scale=0.45,clip,viewport=0pt 0pt 500pt
450pt]{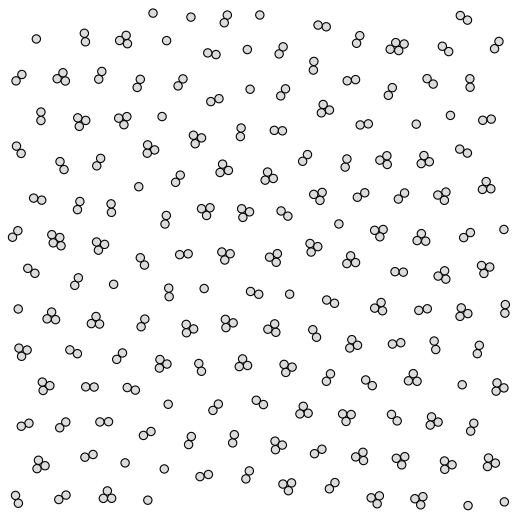}
\caption{\label{fig:simplex-mc} 351 particle MC results for simplex
potential annealed to $k_BT=0.1$.}
\end{figure}

If it is possible to stabilize structures that were once thought to
require directional bonding, what else can be stabilized?  Bilayers?
Block copolymers?  For the purpose of seeing how far isotropic
potentials can be taken, we examined what we call the `five-finger
potential', shown in Fig. \ref{fig:5-finger}.  We chose this form
for the potential since it would inhibit second nearest neighbors of
any lattice to form, only allowing long chains of particles.  Annealed
MC results for this are shown in Fig. \ref{fig:5-finger-mc}.  As
shown, this potential allows for the assembly of such parallel chains
at $\alpha=6.0$.  This potential cannot be built in the lab with
current technology -- it is far too complex, but it shows that
isotropic potentials have perhaps more flexibility than one would
immediately think.  It is also possible that a much simpler potential
could allow for a similar structure to assemble.  

We have also devised a circularly symmetric potential function that
favors the assembly of small clusters of particles.  The form of the
potential was chosen to inhibit the formation of clusters with second
and third (and so on) nearest-neighbors -- the most favorable
structure thus being a {\it simplex}, or equilateral-triangle cluster.
The structures we find are similar to those observed experimentally by
Manoharan et. al. \cite{Pine03}.  Although they do not find the
functional form of the potential explicitly, they conclude that the
clusters that they observe cannot be a result only of van der Waals
attraction and the hard core repulsion of the polystyrene colloid
particles used in their experiment.  This potential is shown in Fig.
\ref{fig:simplex}, which we run at specific area $\alpha=9.6$.  MC
results are shown in Fig. \ref{fig:simplex-mc}.  

One can imagine
carrying on the process of qualitatively searching for isotropic
potentials for more and more complex structures ad infinitum, with
arbitrarily complex structures requiring more and more elaborate
functional forms.  For example, one might try to assemble a Buckyball
with a spherically symmetric pair potential by running an NVT
annealing simulation with 60 particles interacting via a potential
that has sharp minima at every interparticle distance for that
molecule.  There must be a limit, however -- although it is
conceivable that a chiral structure would self-assemble, we cannot
choose its chirality if we employ only an isotropic potential (left
and right handed structures are equally likely).  A key question that
we ask, and that this study answers only in part, is whether we can
make qualitative statements about the types of structures that can be
assembled using only isotropic pair interactions.
  
We are currently working on expanding this work to three dimensional
systems.  Moreover, we are exploring the possibility of tailoring for
self-assembly thermodynamic quantities besides the area or volume,
such as the pressure (in an NPT ensemble simulation) and the chemical
potential (in a $\mu$VT ensemble simulation).  In future work, we plan
to explore the self-assembly properties of multicompontent media using
our inverse/optimization approach.

\begin{acknowledgments}
This work was supported by the Office of Basic Energy Sciences, DOE,
under Grant No. DE-FG02-04ER46108.
\end{acknowledgments}

\end{document}